\documentstyle[fleqn]{tp}

\input psfig

\def\hmpc{\rm \,h^{-1}\,Mpc}

\def\xr{$\xi(r)$\ }

\def\xir{$\xi(r)$\ }
\def\xis{$\xi(s)$\ }
\def\xip{$\xi(r_p,\pi)$\ }
\def\xip{$\xi(r_p,\pi)$\ }
\def\wp{$w_p(r_p)$\ }

\def\kms{\,{\rm km\,s^{-1}}}

\def\ro{r_\circ}
\def\n_med{{\left<n\right>}}

\def\begc{\begin{center} }
\def\endc{\end{center} } 
\def\begf{\begin{figure} }
\def\endf{\end{figure} }

\def\j3{{J_3}}

\def\kms{\mbox{km s$^{-1}$}}

\begin{document}

\twocolumn[
\title{Clustering in the ESP Redshift Survey}
\author{
L. Guzzo$^1$, J.G. Bartlett$^2$, A. Cappi$^3$, S. Maurogordato$^4$, 
E. Zucca$^{3,5}$,\\
G. Zamorani$^{3,5}$, C. Balkowski$^6$, A. Blanchard$^2$, V. Cayatte$^6$, 
G. Chincarini$^{1,7}$,\\
C.A. Collins$^8$, D. Maccagni$^9$, H. MacGillivray$^{10}$, R. Merighi$^3$, 
M. Mignoli$^3$,\\
D. Proust$^6$, M. Ramella$^{11}$, R. Scaramella$^{12}$, G.M. Stirpe$^3$, 
G. Vettolani$^5$\\
{\it $^1$Osservatorio Astronomico di Brera, I-23807 Merate}\\
{\it $^2$Universit\'e L. Pasteur, Obs. Astronomique, F-67000 Strasbourg}\\
{\it $^3$Osservatorio Astronomico di Bologna, I-40126 Bologna}\\
{\it $^4$CERGA, Observatoire de la C\^ote d'Azur, 06304 Nice Cedex 4}\\
{\it $^5$Istituto di Radioastronomia del CNR, I-40129 Bologna}\\
{\it $^6$Observatoire de Paris, DAEC, F-92195 Meudon}\\
{\it $^7$Universit\`a degli Studi di Milano, I-20133 Milano}\\
{\it $^8$Liverpool John-Moores University, Liverpool L3 3AF}\\
{\it $^9$Istituto di Fisica Cosmica e Tecnologie Relative, I-20133 Milano}\\
{\it $^{10}$Royal Observatory Edinburgh, Edinburgh EH9 3HJ}\\
{\it $^{11}$Osservatorio Astronomico di Trieste, I-34131 Trieste}\\
{\it $^{12}$Osservatorio Astronomico di Roma, I-00040 Monteporzio C.}}
\vspace*{16pt}   

ABSTRACT.\
We discuss the two-point correlation properties of galaxies in the 
ESO Slice Project (ESP) redshift survey, both in redshift and
real space.  The redshift-space correlation function \xis for the 
whole magnitude-limited survey is well 
described by a power law with $\gamma \sim 1.55$ between 3 and $\sim 40\hmpc$,
where it smoothly breaks down, crossing the zero value on scales as large 
as $\sim 80\hmpc$.  On smaller scales ($0.2-2\hmpc$), the slope is
shallower mostly due to redshift-space depression by virialized structures,
which are found to be enhanced by the J3 optimal-weighting technique.
We explicitly evidence these effects by computing \xip and the projected
function \wp.  In this way we recover the real-space correlation function 
\xir, 
which we fit below $10\hmpc$ with a power-law model.  This gives a 
reasonable fit, with $r_o=4.15^{+0.20}_{-0.21} h^{-1}\,$ Mpc and 
$\gamma=1.67^{+0.07}_{-0.09}$.  This results on \xir and \xis and
the comparison with other surveys clearly confirm how the shape of
spatial correlations above $\sim 3\hmpc$ is characterised by 
a significant `shoulder' with respect to the small-scale $\sim -1.8$ 
power law, corresponding to a steepening of P(k) near the turnover.

\endabstract]

\markboth{Luigi Guzzo et al.}{Clustering in the ESP Survey}

\small

\section{Introduction}
One of original goals of the ESO Slice Project (ESP) redshift survey
was to study the clustering of galaxies from a survey hopefully not dominated 
by a single major superstructure (as it was the case for the surveys available
at the beginning of this decade), being able to gather sufficient signal 
in the weak clustering regime, i.e. on scales above $10\hmpc$.
After completion of the redshift survey, our first analyses 
concentrated on the galaxy luminosity function (the other main
original goal), for which the ESP has yielded an estimate
with unprecedented dynamic range (Zucca et al. 1997, Z97 hereafter).  
Here we shall report on the more recent results we have obtained on 
the clustering of ESP galaxies.

The ESP covers a strip of sky $1^\circ$ 
thick (DEC) by $\sim30^\circ$ long (RA) (with an unobserved $5^\circ$ sector
inside this strip), in the SGP region.  The target galaxies were selected 
from the EDSGC (Heydon-Dumbleton et al. 1989), and the final catalogue 
contains 3342 redshifts,
corresponding to a completeness of 85\% at a magnitude limit $b_J=19.4$.
More details can be found in Vettolani et al. (1998).  
For the present analysis, we use comoving distances computed within a 
model with $H_o=100\kms$ Mpc$^{-1}$ and $q_o=0.5$, while magnitudes are 
K-corrected as described in Z97.

\section{Redshift-Space Correlations}
%
%
%
\begin{figure*}
\centering\mbox{\psfig{figure=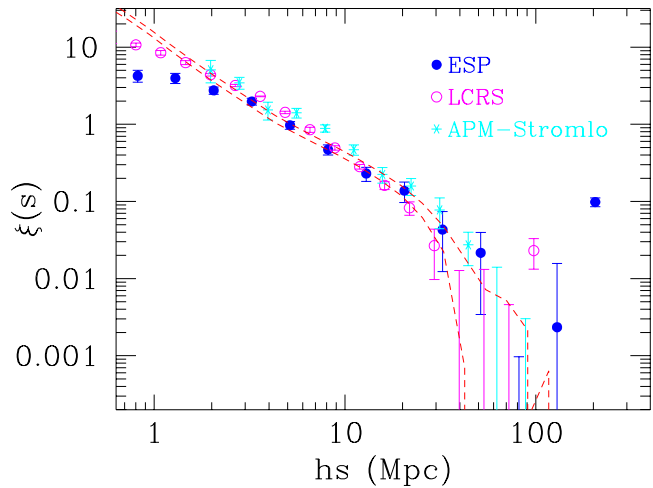,height=7.cm,angle=270}
}
\caption[]{The redshift-space correlation function from the whole ESP
magnitude-limited catalogue, compared to \xis from the LCRS and APM-Stromlo 
surveys.  The dashed lines show the {\sl real space} correlation function
\xir derived from the APM $w(\theta)$, for two different clustering 
evolution models.
}
\label{xi_mlim_surv}
\end{figure*}
\subsection{Optimal Weighting}
The filled circles in Figure \ref{xi_mlim_surv} show \xis computed from
the whole ESP magnitude-limited sample using the
J3-weighting technique (Guzzo et al. 1998, G98 hereafter).  
Note the smooth decay
from the power law at large separations, with correlations going to
zero only around $80\hmpc$, and perhaps some evidence for a positive 
fluctuation around
$\sim 200\hmpc$.  For $r< 3\hmpc$, \xis tends to flatten (see
below for more discussion on this). 
In the same figure, we reproduce \xis 
from the Las Campanas (LCRS, Tucker et al. 1997), and APM-Stromlo 
(Loveday et al. 1992),
redshift surveys.  There is a rather good agreement of the three independent
data sets between 2 and $20\hmpc$, where $\xi(s)=(s/s_\circ)^{-\gamma}$, with
$s_\circ\sim 6 \hmpc$ and $\gamma\sim 1.5$, with perhaps a hint for more power
on larger scales in the 
blue-selected ESP and APM-Stromlo.  In addition, the dashed lines describe
the real-space \xir obtained by de-projecting the angular correlation
function $w(\theta)$ of the APM Galaxy Catalogue (Baugh 1996), for two 
different assumptions on the evolution of clustering.  It is rather 
interesting to note the degree of unanimity (within the error bars),
between the angular and redshift data concerning the large-scale shape 
and zero-crossing scale ($40-90\hmpc$) of galaxy correlations.
In fact, if one ideally extrapolates to larger scales the $\sim -1.8$ 
slope observed in real space  below $3-4\hmpc$ [e.g. from the APM \xir], 
all surveys agree in being consistently above this extrapolation, 
displaying what has been called a `shoulder' or 
a `bump' before breaking down (e.g. Guzzo et al. 1991).
This excess power (see Peacock, these proceedings),
requires in Fourier space a rather steep slope ($\sim k^{-2}$) 
for the power spectrum P(k).  Such a feature is for example common
in CDM models with an additional hot component (Bonometto
\& Pierpaoli 1998), or with high baryonic content (Eisenstein
et al. 1998).

On a different ground, the small amplitude difference between the redshift- 
and real-space correlations on scales $>5\hmpc$ is also remarkable, because
implies a small amplification of \xis due to streaming flows, suggestive
of a low value of $\beta=\Omega_o/b$.  

\subsection{Volume-limited estimates}\label{vol-lim}

\begin{figure*}
\centering\mbox{\psfig{figure=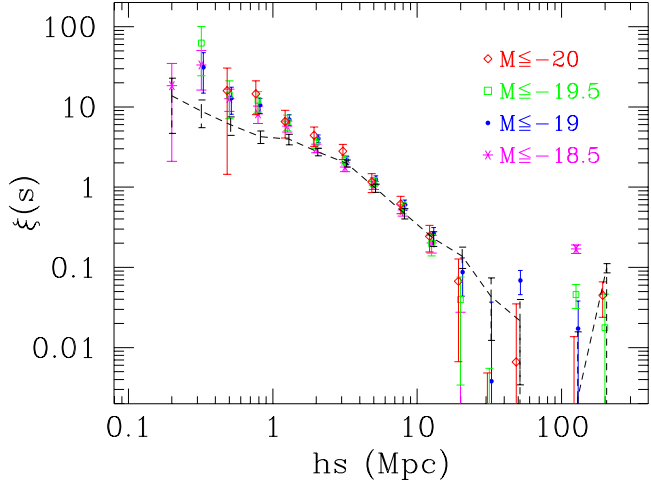,height=7.cm,angle=270}
}
\caption[]{Redshift space correlation function for four volume-limited
subsamples of the ESP survey, compared (dashed line with error bars) to
the J3-weighted optimal estimate from the whole sample.
For clarity, points for $M\le-20$ and $M\le-19$ are displayed with a constant
shift in log(s). 
}
\label{xi_multi}
\end{figure*}
While the J3-weighted estimate has the advantage of maximising the
information extracted from the available data, it has a few drawbacks.
The main one is that of mixing galaxy luminosities in the estimate of
\xis.  The worst aspect of this mixing is that it depends on scale:
in fact, by its own definition, the method weighs pairs depending both 
on their separation and on their distance from the observer (see G98
for details).  
\begin{figure*}
\centering\mbox{\psfig{figure=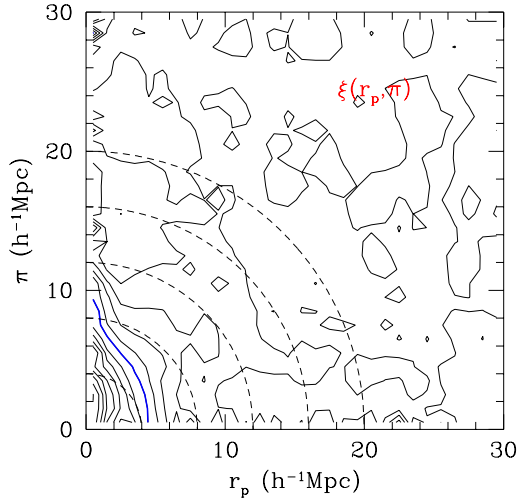,height=7.cm}
}
\caption[]{The \xip map for the whole ESP survey, compared to the  
isotropic correlations expected in the absence of peculiar
velocities (dashed).  The heavy contour corresponds to $\xi=1$.
}
\label{csipz}
\end{figure*}
As a consequence, the main contribution to small-scale correlations 
comes from low-luminosity pairs, that are numerous in the nearer part 
of the sample, while \xis on large scales is mostly the result of 
correlations among high-luminosity objects, that are the in fact
the only ones visible out to the more distant regions of the sample.
Clearly, if there is any luminosity dependence of clustering, 
this technique will modify in some way the shape of \xis.  In particular,
if luminous galaxies are more clustered than faint ones, the 
J3 weighting will tend to produce a {\sl flatter} estimate of \xis.  
In some way, therefore, while this method is certainly optimal for 
maximising the clustering signal on large scales (and so the previous 
discussion is certainly valid), it might be dangerous to
draw from it conclusions on the {\sl global} shape of \xis from, say,
$0.1$ to $100 \hmpc$.   A wise way to counter-check the results 
produced using the optimal weighting technique, 
is that of estimating \xis also from volume-limited subsamples extracted
from the survey.  In this case, each sample includes a narrower range
of luminosities, and no weighting is required, the density of objects 
being the same everywhere.  We have performed this exercise on the
ESP, selecting four volume-limited samples.  The result is quite 
illuminating and is shown in Figure~\ref{xi_multi}, directly compared
to the J3-weighted estimate (dashed line).  One can see that in fact
at least in the case of our data, the latter method 
produces a \xis which is shallower below $\sim 3\hmpc$.  Note, however,
how on larger scales the shape is consistent between the two methods, 
and for $r>20\hmpc$ the J3 technique performs much better in terms of 
signal-to-noise, than the single volume-limited samples.   By studying 
the distortions of \xip and its projections (see below), we have seen
that this small-scale redshift-space shape of the J3 estimate
arises from a combination of a smaller real-space clustering and
a larger pairwise velocity dispersion (see G98).  

Finally, from this figure one cannot notice any explicit dependence
of clustering on luminosity, within the magnitude range considered.
We shall see in the next section how this will change when we explore
correlations free of redshift-space distortions.

\section{\xip and Real-Space \\ Correlations}
More can be learned from redshift survey data, if we are able to disentangle 
the effect of clustering from that of the small-scale dynamics of galaxies,
which adds to the Hubble flow to produce the observed redshift.
This is traditionally done through the function \xip (see e.g. G98), 
that essentially describes galaxy correlations as a function
of two variables, one perpendicular ($r_p$), and the other parallel ($\pi$),
to a sort of mean line of sight defined for each galaxy pair.  
In Figure~\ref{csipz}, we show \xip computed for the whole ESP survey, using
the same technique used for $\xi(s)$.  From this figure, one can clearly see 
the small-scale stretching of the contours along $\pi$, produced by the 
relative velocity dispersion of pairs within clusters and groups.

Projecting \xip onto the $r_p$ axis, one gets
%
%
%
\begin{eqnarray}
w_p(r_p) &\equiv& 2 \int_0^{\infty} dy\, \xi(r_p,\pi)\nonumber \\ 
&=& 2 \int_0^{\infty} dy \, \xi\left(r\right)\, ,\label{wp}
\end{eqnarray}
%
%
where now \xr is the real-space correlation function,
with $r=\sqrt{r_p^2 + y^2}$.   \wp is therefore independent from the 
redshift-distortion field,
and is analytically integrable in the case of a power-law 
$\xi(r)= (r/r_0)^{-\gamma}$.  Given this form, the values of 
$r_o$ and $\gamma$ that best reproduce the data can be evaluated
through an appropriate best-fitting procedure (G98).
By applying this to the
map of Figure~\ref{csipz}, we recover 
$r_o=4.15^{+0.20}_{-0.21} h^{-1}\,$ Mpc and $\gamma=1.67^{+0.07}_{-0.09}$.
As we anticipated in the previous section, this value of $r_o$ is slightly 
smaller than the value $\sim 5 \hmpc$ which is measured by most other 
surveys, as e.g. the LCRS, and is an indication that our  
J3-weighted estimate on small scales could be biased in a subtle way towards
faint, less clustered galaxies.  In fact, while on one side the weighting 
scheme certainly amplifies the small-scale pairwise dispersion 
($\sigma_v(1)\simeq 650\; \kms$ with respect to $\sim 380\; \kms$ measured
from the $M\le-20$ volume-limited sample), at the same time 
the contribution in projection from cluster galaxies seems to be still 
relatively low. This is in fact reasonable, given the small
volume of the ESP ($\sim 1.9 \cdot 10^5{\rm \,h^{-3}\,Mpc^{3}}$ at 
the effective depth of the survey.)
\begin{figure*}
\centering\mbox{\psfig{figure=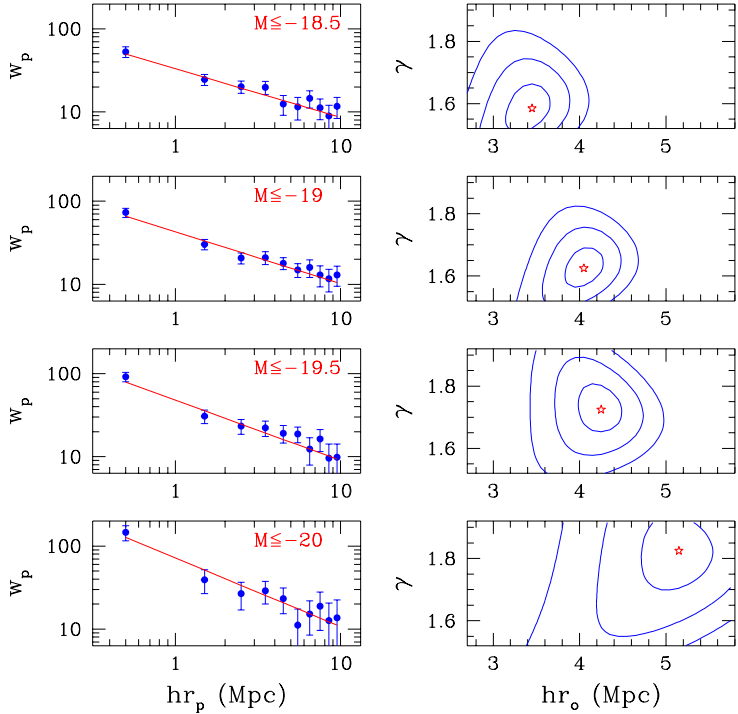,height=10.cm}
}
\caption[]{Best-fit results of a power-law \xir to the projected
function \wp for the four volume-limited samples.
}
\label{wpfit_vlim}
\end{figure*}

Finally, we have computed \xip and \wp for the four 
volume-limited subsamples introduced in \S~\ref{vol-lim}.  Figure \ref{wpfit_vlim} shows the 
result of the power-law fits to \wp.  A weak, but 
significant growth of $\ro$ and $\gamma$ with luminosity,
especially for $M<M^*\simeq-19.5$, is evident.  This is in contrast 
with the conclusions one could have drawn from the behaviour of \xis 
in Figure~\ref{xi_multi}.  The reason for this lays in the 
growing amount of small-scale redshift-space distortion in the four samples, 
that counter-balances the growth of clustering.

%



\begin{thebibliography}{99}

\bibitem{Bono} Bonometto, S.A., \& Pierpaoli, E., 1998, NA, 3, 391
\bibitem{Baugh} Baugh, C.M., 1996, MNRAS, 280, 267
\bibitem{F94a} Fisher, K.B., Davis, M., Strauss, M.A., 
   Yahil, A., \& Huchra, J.P. 1994a, MNRAS, 266, 50 (F94a)
\bibitem{}Eisenstein, D., Hu, W., Silk, J., \& Szalay, A.S., 1998, ApJ, 494, L1
\bibitem{Guz97} Guzzo, L., Strauss, M.A., Fisher, K.B., Giovanelli, R., 
and Haynes, M.P. 1997, ApJ,, 489, 37
\bibitem{Guz98} Guzzo, L., \& the ESP Team, 1998, A\&A,, submitted
\bibitem{Hey} Heydon-Dumbleton, N.H., Collins, C.A., MacGillivray, H.T., 1989,
           MNRAS, 238, 379
\bibitem{Lov} Loveday, J., Efstathiou, G., Peterson, B. A. Maddox, S. J., 
1992, ApJ, 400, 43L 
\bibitem{Tuc} Tucker, D.L., Oemler, A., Kirshner, R.P., et al., 1997,
MNRAS, 285, L5
\bibitem{Vet97} Vettolani, G., \& the ESP Team, 1997, A\&A, 325, 954
\bibitem{Vet98} Vettolani, G., \& the ESP Team, 1998, A\&AS, 130, 323
\bibitem{Zuc97} Zucca, E., \& the ESP Team, 1997, A\&A, 326, 477

\end{thebibliography}
\end{document}